# THE STARRY UNIVERSE OF JOHANNES KEPLER

Christopher M. Graney

Jefferson Community & Technical College

1000 Community College Drive

Louisville, Kentucky 40272 (USA)

christopher.graney@kctcs.edu

ABSTRACT

Johannes Kepler described the Copernican universe as consisting of a central, small, brilliant sun with its planetary system, all surrounded by giant stars. These stars were far larger than, and much dimmer than, the sun—his *De Stella Nova* shows that every visible star must exceed the size of the Earth's orbit, and the most prominent stars may exceed the size of the entire planetary system. His other writings, including his response to Ingoli, his *Dissertatio cum Nuncio Sidereo*, and his *Epitome Astronomiae Copernicanae*, also reflect this Copernican universe. To Kepler, such a universe was an illustration of divine power—and solid evidence against the stars being suns, against the universe of Giordano Bruno. Kepler's starry universe was in fact the Copernican universe supported by observations of the stars, which showed them to have measureable apparent sizes. Not until the later seventeenth century were those apparent sizes shown to be spurious, allowing for a universe in which the stars were suns.

INTRODUCTION

Robert Hooke presented an observation in his 1674 *An Attempt to Prove the Motion of the Earth* that was more significant than his supposed observation of annual parallax that was the central feature of the book. Hooke discussed how he had been able to observe through his telescope during the daylight a star at the zenith, and how that star appeared to be very, very small:

> [B]y this Observation of the Star in the day time when the Sun shined, with my 36 foot Glass I found the body of the Star so very small, that it was but some few thirds [of an arc second] in Diameter, all the spurious rayes that do beard it in the night being cleerly shaved away, and the naked body thereof left a very small white point.
>
> The smalness of this body thus discovered does very fully answer a grand objection alledged by divers of the great *Anti-copernicans* with great vehemency and insulting; amonst which we may reckon *Ricciolus* and *Tacquet*, who would fain make the apparent Diameters of the Stars so big, as that the body of the Star should contain the great Orb [Earth's orbit] many times, which would indeed swell the Stars to a magnitude vastly bigger then the Sun, thereby hoping to make it seem so improbable, as to be rejected by all parties. But they that shall by this means examine the Diameter of the fixt Stars, will find them so very small that according to these distances and Parallax they will not much differ in magnitude from the body of the Sun....[1]

The objection that Hooke notes here, of star size against the Copernican hypothesis, dated back to Tycho Brahe in the late sixteenth century. It was a simple argument—that for stars both to have the apparent size that they do, and also to be as distant as required by the Copernican hypothesis, every last one of them must be enormous. Thus the Copernican hypothesis essentially turned the stars into a new class of giant celestial bodies, far different than the earth, sun, moon, and planets, and this was so improbable as to prompt the rejection of the Copernican hypothesis.

What Hooke did not note was that many Copernicans embraced the giant stars. In fact, one Copernican who embraced them enthusiastically was an astronomer who Hooke cited several times in *An Attempt to Prove*, Johannes Kepler.[2] Indeed, Kepler used the star sizes to argue against a sort of universe that Hooke alluded to in *An Attempt to Prove*: a universe of "almost

C. M. Graney: The Starry Universe of Johannes Kepler, pre-print (page 2 of 28)

infinite extension"; a universe in which "all the fixt Stars" are "as so many Suns"; a universe like that of Giordano Bruno.³ Kepler embraced the giant stars in exactly the sort of manner that prompted vehement complaints and insults from anti-Copernicans like Giovanni Battista Riccioli: as not only a proof that the universe is not universe of suns, but as an illustration of the power and wisdom of God.

GIANT STARS

Under the Copernican heliocentric hypothesis the Earth's orbit must be negligible in size compared to the distances to the stars lest Earth's annual motion yield detectable parallax in the stars. Brahe noted that stars have a measurable apparent size, determining that first magnitude stars measured a little less than one tenth of the apparent diameter of the moon (a little less than three minutes of arc); his measurements were generally consistent with those of Ptolemy and others.⁴ Since an object of a given apparent size has a physical size that is some fixed fraction of its distance (Figure FS), at the vast distances required for the stars in the heliocentric hypothesis, Brahe's measured apparent sizes translated into enormous physical bulks: every star must dwarf the sun, which in turn must be a unique, small body in a universe of giants.

Johann Georg Locher and his mentor Christoph Scheiner neatly summarized Brahe's giant stars objection in their 1614 book *Disquisitiones Mathematicae*. They wrote that in the Copernican hypothesis the Earth's orbit is like a point within the universe of stars; but the stars, having measurable sizes, are larger than points; therefore, in the Copernican hypothesis every star must be larger than Earth's orbit, and of course vastly larger than the sun itself:

> [E]ven the smallest star visible to the eye is much larger than the whole circle of Earth's orbit. This is because such a star has a measurable size, as does the circumference of the sky. The ratio of the size of the star to the size of the firmament of fixed stars is therefore perceptible. But according to the Copernican opinion, the ratio of the size of the circle of Earth's orbit to the size of the firmament is imperceptible. For in the Copernican opinion the size of the Earth's orbital circle holds the same proportion to the firmament as the size of Earth itself holds to the firmament in the common geocentric opinion. Yet experience shows the Earth to be of imperceptible size compared to the firmament. Thus in the Copernican opinion it is the circle of Earth's orbit that is of imperceptible size compared to the firmament—and therefore smaller than the smallest perceptible star.[5]

The giant stars of the Copernican hypothesis stood in contrast to the more commensurate star sizes found in Brahe's own hybrid geocentric (or geo-heliocentric) hypothesis. In it, the sun, moon, and stars circled an immobile Earth, but the planets circled the sun (Figure TB). Insofar as these bodies were concerned, Brahe's hypothesis was observationally and mathematically identical to the Copernican hypothesis, and thus fully compatible with telescopic discoveries regarding these bodies. But the fixed stars were a different matter. Since the Earth was immobile in Brahe's hypothesis, there was no expectation of parallax, and thus no need for the stars to be distant in order to explain the absence of detectable parallax. Brahe put the stars a bit beyond Saturn, at a distance from the sun of 14,000 times the semidiameter of the Earth. And, since the stars were roughly similar to Saturn both in terms of distance and in terms of appearance in the night sky, they had to be similar to Saturn in physical bulk, too. In the Tychonic hypothesis, the sizes of the Earth, sun, moon, and planets were commensurate, with the moon being smallest and the sun being largest. This stood in contrast to the Copernican hypothesis, where according to Brahe's measurements and calculations, every last star dwarfed sun, moon, and planets (Figure TS).

The giant stars were, according to Christian Huygens, Brahe's "principle argument" against Copernicus.[6] Brahe considered them an absurdity. And, as Albert Van Helden has written, "Tycho's logic was impeccable; his measurements above reproach. A Copernican simply had to accept the results of this argument" and agree that the stars were giant.[7] Following Brahe a number of people raised the star size issue against the Copernicans. These included Simon Marius, who in 1614 noted that stars seen with the telescope still had circular shapes and measurable sizes, and who cited this in favor of the Tychonic hypothesis; Locher and Scheiner, also in 1614 as mentioned previously; Francesco Ingoli, who raised the star size issue in a 1616 essay to Galileo that Galileo believed to have been influential in the rejection of the Copernican hypothesis by the Congregation of the Index in Rome later that year; Marin Mersenne, who highlighted the star size question while contrasting the Tychonic and Copernican systems in his 1627 *Traité de l'harmonie universelle*; Peter Crüger, who in his 1631 *Cupediae Astrosophicae* remarked that, granted the star size question, "I therefore do not understand how the Pythagorean or Copernican world system can survive"; and Riccioli, who analyzed 126 arguments for and against Copernicus in his 1651 *Almagestum Novum*, and considered star sizes to be one of the few decisive arguments for either side.[8]

As has been previously discussed in the pages of this journal, Riccioli gave an extensive treatment of the star size issue, particularly in light of the telescope.[9] The telescope did not change the issue, because the small aperture telescopes used at the time produced measurable disks when directed toward the fixed stars (Figure D), yet did not yield a detection of parallax. (These disks were not then understood to be spurious products of the diffraction of light waves—the "Airy disk" phenomenon.[10]) The telescopic disks were smaller than what Brahe had measured, and this was attributed to the telescope stripping away glare and revealing the true bodies of the fixed stars, much as Hooke described above (and just as the telescope revealed the true bodies of wandering stars, or planets[11]). Riccioli showed that the telescopically measured star sizes, combined with the increased ability to detect parallax that the telescope provided, still yielded giant stars. As Locher and Scheiner had noted, so long as the apparent sizes of stars were measurable, but annual parallax was not, then every last star had to be larger than Earth's orbit.

Both Riccioli and Locher-Scheiner complained that Copernicans just waved away this problem by appealing to the power of God. "They go on," Locher and Scheiner wrote in *Disquisitiones*, "about how from this everyone may better perceive the majesty of the Creator," and they considered that "laughable."[12] Riccioli rejected invoking the power of God as a response to the star-size argument, stating that "even if this falsehood cannot be refuted, nevertheless it cannot satisfy the more prudent men."[13]

KEPLER'S STARRY UNIVERSE IN *DE STELLA NOVA*

Kepler was one Copernican who both accepted that the stars were giant and viewed them in terms of divine power. This is perhaps most clearly evident in his 1606 *De Stella Nova*. Within Chapter 16 of that book[14] we find him discussing Brahe's view on giant stars, grousing about how—

> Brahe finds a lack of elegance in the most perfect of works, if the vastness of the sphere of one of the fixed stars be so insane; the meagerness of all the wandering stars [planets] so contemptible. How huge the fault in the human body, he says, if the finger, if the nose, might surpass by many times the bulk of the whole remainder of the body.[15]

Kepler counters that such variation in size is not unreasonable. Consider, he says, a 120-foot serpent noted by Pliny, versus a mite: the length of the snake exceeds that of the mite by a factor of 100,000. Kepler also compares the size of human beings to the Earth and to the universe. A variety of sizes clearly exists in the universe.[16] Thus he finds no problem in saying that the distance from the sun to the fixed stars holds the same proportion to the orbit of Saturn as the distance from the sun to Saturn holds to the diameter of the sun itself—and so he estimates the distance from the sun to the fixed stars at just over 34,000,000 times the semidiameter of Earth. The sun seen from Earth has an apparent diameter of thirty arc minutes, and Saturn is ten times farther from the sun than Earth, so the sun seen from Saturn would have an apparent diameter of three minutes. And therefore, says Kepler, the orbit of Saturn seen from the fixed stars would have an apparent diameter of three minutes.[17]

Kepler argues that what is commensurate in a Copernican universe are speeds. "The perfection of the universe is motion," he says.[18] He shows that in a Copernican universe, speeds range from Saturn, moving at 300 German miles per hour, to Mercury, moving at 1000—"a beautiful proportion," he writes, "where what is nearer to the quiescent sun (the dispenser of all movement) is always swifter."[19] Even the speeds of the day and night sides of Earth, and the velocity of the moon, fall into this same general range. Everything in the Copernican solar system moves at speeds ranging from about 250 to about 1250 miles per hour.

Kepler contrasts this with the geocentric universe:

> Go now to Ptolemy and the ancient opinion; you will find everything more incredible. In that, the semidiameter of the sphere of the fixed stars occupies twenty thousand semidiameters of Earth. The circumference therefore will be 63,000—truly a reasonable number,[20] compared to the Copernican, but which all is said to go round in one day. Therefore 2,625 semidiameters (each of which contains 860 miles) are covered in one hour.[21] Behold here what to me is an immense distinction. In the view of Ptolemy, Saturn is the nearest to the fixed stars, such that it will almost touch them. Following

Copernicus, in one hour it traverses 300 miles; following Ptolemy, 2,257,500 miles.[22] Saturn must be believed to be 7,525 times swifter[23] under Ptolemy, than under Copernicus. Whoever attempts mentally to comprehend this incredible velocity is overcome just as much as, and indeed more severely than, someone who attempts to comprehend the Copernican immensity.[24]

Kepler notes that Tycho Brahe's hypothesis yields a somewhat more compact universe than Ptolemy's, and thus somewhat lower speeds, but the geocentric speed problem remains.[25] Kepler adds that it is more credible to have a vast thing with no motion, than a small thing with great motion.[26] Kepler also notes that size means nothing to God, and here his writing becomes almost poetic:

> Where magnitude waxes, there perfection wanes, and nobility follows diminution in bulk. The sphere of the fixed stars according to Copernicus is certainly most large; but it is inert, no motion. The universe of the movables [the planets] is next. Now this—so much smaller, so much more divine—has accepted that so admirable, so well-ordered motion. Nevertheless, that place neither contains animating faculty, nor does it reason, nor does it run about. It goes, provided that it is moved. It has not developed, but it retains that impressed to it from the beginning. What it is not, it will never be. What it is, is not made by it—the same endures, as was built. Then comes this our little ball, the little cottage of us all, which we call the Earth: the womb of the growing, herself fashioned by a certain internal faculty. The architect of marvelous work, she kindles daily so many little living things from herself—plants, fishes, insects—as she easily may scorn the rest of the bulk in view of this her nobility. Lastly behold if you will the little bodies which we call the animals. What smaller than these is able to be imagined in comparison to the universe? But there now behold feeling, and voluntary motions—an infinite architecture of bodies. Behold if you will, among those, these fine bits of dust, which are called Men; to whom the Creator has granted such, that in a certain way they may beget themselves, clothe themselves, arm themselves, teach themselves an infinity of arts, and daily accomplish the good; in whom is the image of God; who are, in a certain way, lords of the whole bulk. And what is it to us, that the body of the universe has for itself a great breadth, while the soul lacks for one? We may learn well therefore the pleasure of the Creator, who is author both of the roughness of the large masses, and of the perfection of the smalls. Yet he glories not in bulk, but ennobles those that he has wished to be small.

In the end, through these intervals from Earth to the Sun, from Sun to Saturn, from Saturn to the fixed stars, we may learn gradually to ascend toward recognizing the immensity of divine power.[27]

Having prepared the ground with these discussions, Kepler now comes directly to the question of star sizes. Since he has stated that the orbit of Saturn would have an apparent diameter of three seconds as seen from the sphere of the stars, any star with an apparent diameter of three seconds as seen from Earth must be equal in physical size to the orbit of Saturn—that is, to the entire planetary system. And so Sirius, the most prominent of all the stars, which Kepler says to appear larger than three minutes, must be larger than the entire planetary system, and the awesome "new star" or *nova* of 1604 that is the subject of *De Stella Nova* must be larger still:

> I have gladly inserted so much here concerning the objections to the Copernican vastness of the fixed stars, because it all pertains to the incredible magnitude that must be estimated for the new star. For if it occupies only four minutes (the size Sirius appears), then through this hypothesis of Copernicus it is much greater than the whole machinery of the movables [the planetary system]. For earlier we were granting to that machinery only three minutes, were it to be seen from the fixed stars.[28]

It follows from Kepler's numbers that Sirius and the nova must each rival or exceed the size of an entire geocentric universe, since, as Kepler has noted, in geocentric hypotheses the fixed stars lay just beyond Saturn. Furthermore, any star whose physical size would be the same as Earth's orbit, namely one tenth the size of Saturn's orbit, would have an apparent diameter of three tenths of a minute of arc, or 18 seconds of arc (or, to provide a point of comparison to Hooke's star size, 1080 thirds of arc). This is the apparent diameter that Brahe had determined for sixth-magnitude stars, barely visible to the eye. And since according to Kepler the physical diameter of the sun is less than one hundredth that of Earth's orbit, clearly every last star in the sky utterly dwarfs the sun.

In short, Kepler's views are precisely the sort that Locher and Scheiner complain about regarding Copernicans: he believes all the fixed stars—every last visible one—are larger than the Earth's orbit; and he says this is reasonable because of divine power.

KEPLER'S STARRY UNIVERSE IN HIS OTHER WORKS

The views seen in Chapter 16 of *De Stella Nova* also appear in Kepler's other writings. In the response he wrote to Ingoli's 1616 essay to Galileo, Kepler asks, in criticizing Ingoli's assertion that a universe of giant stars is "asymmetrical"—

> [B]ased on what laws might he examine the works of the hands of God, so as to declare them out of proportion? I have shown [in *De Stella Nova*] the proportion to be greater between a mite in the skin of the hand of a man and that [120-foot] African serpent.... Why, in the eyes of Ingoli, is a distance of 16,506,000 semidiameters of Earth [Ingoli's estimated Copernican distance to the stars] excessive, but not 14,000? Men make what comparison? Based on what human examples might the confident mind of Ingoli reject the works of God as excessive?[29]

And then, responding to a vague suggestion by Ingoli that perhaps the star size problem could be avoided if somehow the fixed stars did not "operate" the same as other celestial objects[30] (a suggestion that seems to have been a lucky guess—there is no reason to think Ingoli knew anything about diffraction and spurious stellar images):

> Has he said, the fixed stars are unable to *operate* on Earth? We may say nothing concerning *operating*, a thing not acknowledged by all: we may say something concerning *illumination*, which is the operation which lies open to the eyes. Why might those fixed stars, which illuminate through 14,000 semidiameters, not illuminate through 16,506,000? If they are a thousand times more remote, they will also be that many times larger: thus the effect of the illumination of the Earth will remain the same.[31]

Thus Kepler expects that a thousand-fold increase in the distance of the stars means a thousand-fold increase in their physical size as well, per Figure FS. Giant stars are simply part of his view of the Copernican universe. Referring to *De Stella Nova*, he notes, "I have dissolved the pretended absurdity of the magnitudes of the fixed stars"[32] in the Copernican hypothesis. But it is only the *absurdity* that Kepler claims to dissolve in his response to Ingoli, not the magnitudes themselves; not the giantness of the stars. That giantness is, as Van Helden said, the logical result of the measured apparent sizes of the stars and of the vast distance to them required by the Copernican hypothesis.

The sizes of the stars also appear when Kepler writes against a Bruno-style universe, in which all the stars are suns. Alexandre Koyré treated Kepler's views on this matter in his 1957 work *From the Closed World to the Infinite Universe*,[33] emphasizing the correctness of Kepler's views given the knowledge available at the time. Koyré noted,

> [I]f we have to restrict its [the universe's] contents to the visible stars, which moreover appear to us as finite, measurable bodies—not points of light—we will never be able to assign to them a uniform distribution that would "save" the phenomena....[34]

Kepler's view of the Copernican universe, with its giant stars and small sun—and not stars that are suns—was what measurement and logic demanded, and Koyré repeatedly emphasized that Kepler's views were driven by observations. As the sun logically is not one of the stars, but much smaller, then stars when seen from each other must look much larger than they do when seen from Earth.[35] Kepler writes in Chapter 21 of *De Stella Nova*,

> Indeed, let us take, for instance, three stars of the second magnitude in the belt of Orion, distant from each other by 81′, being, each one, of at least 2 minutes in diameter. Thus, if they were placed on the same spherical surface of which we are the center, the eye located on one of them would see the other as having the angular magnitude of about 2¾°; [a magnitude][36] that for us on the earth would not be occupied by five suns placed in line and touching each other. And yet these fixed stars are by no means those that are the nearest to each other; for there are innumerable smaller ones that are interspersed [between them]. Thus if somebody were placed in this belt of Orion, having our sun and the center of the world above him, he would see, first, on the horizon, a kind of unbroken sea of immense stars quasi-touching each other, at least to the sight; and from there, the more he raised his eyes, the fewer stars would he see; moreover, the stars will no longer be in contact, but will gradually [appear to be] more rare and more dispersed; and looking straight upward he will see the same [stars] as we see, but twice as small and twice as near to each other.[37]

Moving the stars off a sphere does not change the situation, because the farther any star is removed from Earth, the larger it must be[38] (again, per Figure FS):

> For if the thing is as thus said, then certainly when any one [of the Orion belt stars] will be higher by two, three, a hundred times, it will therefore be larger by two, three, a hundred times. Indeed, you may say it is elevated however much; you will never arrange things so that it may be seen by us to not have a diameter of two minutes.[39]

Thus the stars are large, but Kepler notes elsewhere in writing against Bruno that they are not very bright—all the more indication that the universe is not a universe of suns. In his 1610 *Dissertatio cum Nuncio Sidereo*, he writes:

> [Let Bruno] not lead us on to his belief in infinite worlds, as numerous as the fixed stars and all similar to our own. Your [Galileo's] third observation comes to our support: the countless host of fixed stars exceeds what was known in antiquity. You do not hesitate to declare that there are visible over 10,000 stars. The more there are, and the more crowded they are, the stronger becomes my argument against the infinity of the universe, as set forth in my book on the "New Star," Chapter 21, page 104. This argument proves that where we mortals dwell, in the company of the sun and the planets, is the primary bosom of the universe; from none of the fixed stars can such a view of the universe be obtained as is possible from our earth or even from the sun. For the sake of brevity, I forbear to summarize the passage. Whoever reads it in its entirety will be inclined to assent.

> Let me add this consideration to buttress my case. To my weak eyes, any of the larger stars, such as Sirius, if I take its flashing rays into account, seems to be only a little smaller than the diameter of the moon. But persons with unimpaired vision, using astronomical instruments that are not deceived by these wavy crowns, as is the naked eye, ascertain the dimensions of the stars' diameters in terms of minutes and fractions of minutes. Suppose that we took only l000 fixed stars, none of them larger than l′ (yet the majority in the catalogues are larger). If these were all merged in a single round surface, they would equal (and even surpass) the diameter of the sun. If the little disks of 10,000 stars are fused into one, how much more will their visible size exceed the apparent disk of the sun? If this is true, and if they are suns having the same nature as our sun, why do not these suns collectively outdistance our sun in brilliance? Why do they all together transmit so dim a light to the most accessible places? When sunlight bursts into a sealed room through a hole made with a tiny pin point, it outshines the fixed stars at once. The difference is practically infinite; if the whole room were removed, how great would it become? Will my opponent tell me that the stars are very far away from us? This does not help his cause at all. *For the greater their distance, the more does every single one of them outstrip the sun in diameter.*[40]

Thus the stars are huge compared to the sun, but their light is weak. This is even though Kepler says that, like the sun, they "generate their light from within."[41] "It is quite clear," he writes, "that the body of our sun is brighter beyond measure than all the fixed stars together, and therefore this world of ours does not belong to an undifferentiated swarm of countless others."[42]

Kepler revisits these same ideas early in his *Epitome Astronomiae Copernicanae*. Here again is the discussion of the stars in Orion's belt.[43] Here again is the discussion of how the farther away a star is supposed to be, the larger its true physical size must be.[44] Here again is the use of star sizes as an argument against Bruno.[45] And *Epitome* contains something more—a diagram of what the universe of stars might look like (Figure EP), showing a small sun surrounded by much larger stars.[46] Given all this, there is a surprising anomaly several hundred pages later in the *Epitome*.[47] Here Kepler contradicts his earlier discussion by stating that stars are incredibly small—just a tiny fraction of the semidiameter of the sun. This stems from an argument about the nature of the universe that is based on assumptions Kepler makes regarding matter and the densities of different regions of the universe, not on observations. Kepler justifies these tiny stars by means of a general statement that the fixed stars, when observed by a skilled astronomer using a telescope, appear as mere points of light. However, mere points go against everything else Kepler has said about stars. While this later stars discussion retains the idea that the sun is unique (because the stars are now so tiny versus the sun, rather than so large), stars that are points—that is, immeasurably small—no longer scale in size with distance per Figure FS. Nothing can be said about their surface brightness. Thus, they can in fact be Bruno's suns at vast distances. But the fact is, one telescopic astronomer after another in the first two thirds of the seventeenth century referred to stars seen through telescopes as being finite, measurable disks, not as being immeasurable points: Galileo, Marius, Hortensius, Riccioli, even Hevelius.[48] Kepler's "points" comment—like Ingoli's comment about stars and "operating" that Kepler dismissed—seem anomalous. Kepler's Figure EP and his earlier *Epitome* discussion regarding stars are what agree with what he sketched out in his various other writings, and are what hold up under the observations of that time.

THE COPERNICAN UNIVERSE REQUIRED BY OBSERVATIONS—A SUMMARY

What Kepler sketches out in his various writings is exactly what must be the case, given observations of the visible stars as finite, measurable bodies, and given the Copernican hypothesis. The stars must be vastly distant in order to not reveal any annual parallax. But since they have finite, measureable apparent sizes, then in terms of physical bulks they must be enormous. The early telescope reveals that the apparent sizes of stars might not be as large as Brahe determined with non-telescopic instruments, but it reveals apparent sizes nonetheless (per Figure D), and thus the stars must have physical bulks at least as large as the Earth's orbit, since (as Locher and Scheiner said) Earth's orbit is vanishingly small in a Copernican universe, while the stars are merely small. Since there are so many stars, the total area of the sky that they occupy must be a non-negligible fraction of the area occupied by the disk of the sun; yet their combined power of illumination is negligible compared to the sun. Thus although the stars must be vast in bulk, they also must be dim—they have low surface brightness. The more distant the stars might be, the greater their bulks must be, and thus the lower their surface brightness.

Therefore, in the early seventeenth century, science reveals the Copernican universe to consist of exactly that which Kepler describes in the sixteenth chapter of *De Stella Nova*: a vast shell of huge but dull stars, surrounding a tiny but brilliant sun and its lively planets; and at least one of these planets teems with life; and among that life are these motes of dust called human beings, which beget themselves, clothe themselves, and so forth. A universe of sun-like stars, on the other hand, is the creation of those who don't do their science carefully enough. It is, says Kepler, a universe invented by crazy philosophers, who have broken out of their asylum and are

C. M. Graney: The Starry Universe of Johannes Kepler, pre-print (page 13 of 28)

running around exulting in the immensity they have imagined.[49] It is not the universe observed by careful astronomers.

OTHER COPERNICANS

Since Kepler's view of the Copernican universe is what follows from observations, it is not surprising to find that it was not unique.[50] In 1576 the Copernican Thomas Digges had also used language which spoke of the starry universe in terms of both giant stars and divine power, when he published his English translation of portions of *On the Revolutions* together with a sketch (Figure TD) of the Copernican universe under the heading "A Perfit description of the Cœlestial Orbes". Kepler's sketch from *Epitome* shares some features with it. Within his sketch Digges writes this description of the universe of stars:

> THE PALLACE OF FOELICITYE GARNISHED WITH PERPETUALL SHININGE GLORIOUS LIGHTES INNUMERABLE. FARR EXCELLINGE OUR SONNE BOTH IN QUANTITYE AND QUALITYE THE VERY COURT OF COELESTIALL ANGELLES DEVOYD OF GREEFE AND REPLENISHED WITH PERFITE ENDLESSE IOYE THE HABITACLE FOR THE ELECT

In his translation of Copernicus, he adds the following commentary:

> Herein can we never sufficiently admire this wonderful & incomprehensible huge frame of God's work proponed to our senses.... we may easily consider what little portion of God's frame our Elementary corruptible world is, but never sufficiently of that fixed Orb garnished with lights innumerable and reaching up in Spherical altitude without end.... And this may well be thought of us to be the glorious court of the great God, whose unsearchable works invisible, we partly by these his visible, conjecture; to whose infinite power and majesty, such an infinite place, surmounting all other both in quantity and quality, only is convenient.

Then there is the Copernican Christoph Rothmann, with whom Tycho corresponded. When Tycho put the star size argument to Rothmann, with the assertion that the giant Copernican stars were absurd, Rothmann responded with—

> [W]hat is so absurd about [an average star] having size equal to the whole [orbit of Earth]? What of this is contrary to divine will, or is impossible by divine Nature, or is inadmissible by infinite Nature? These things must be entirely demonstrated by you, if you will wish to infer from here anything of the absurd. These things which vulgar sorts see as absurd at first glance are not easily charged with absurdity, for in fact divine Sapience and Majesty is far greater than they understand. Grant the Vastness of the Universe and the Sizes of the stars to be as great as you like—these will still bear no proportion to the infinite Creator. It reckons that the greater the King, so much more greater and larger the palace befitting his Majesty. So how great a palace do you reckon is fitting to GOD?

And there is Philips Lansbergen's discussion of this matter, found in his 1629 *Bedenckinghen, op den daghelijcksen, ende iaerlijcksen loop vanden aerdt-cloot,* one of the first public defenses of the Copernican system. *Bedenckinghen* was translated into Latin by Hortensius and published as *Commentationes in Motum Terrae Diurnum, et Annuum*. Lansbergen accepted the giant Copernican stars, and wrote a lengthy discussion on them in which he used a variety of passages from Scripture to justify their existence. Indeed, Lansbergen saw the giant stars as the warriors in God's own army, and the palace guard of heaven—and he thanked the telescope for providing the information needed to show that this could be true (namely, the telescope's revelation that far more stars existed than could be seen with the naked eye, and so they were numerous enough to make an army suitable for God).

All of these Copernicans speak of a universe of giant stars (or stars excelling our sun), just like Kepler. Unlike Kepler, however, Lansbergen and Digges, at least, preferred to think of the giant stars as brilliant, not dull. Lansbergen notes that thus they would harmonize with 1 Timothy 6:16, which speaks of God dwelling in inaccessible light.

CONCLUSION

Kepler's universe, with its giant, dull stars surrounding a single tiny, brilliant sun and its planets, is the Copernican universe that astronomical observations required in Kepler's time, and it would remain the Copernican universe until solid, reproducible observations began to suggest that the

C. M. Graney: The Starry Universe of Johannes Kepler, pre-print (page 15 of 28)

sizes of stars, even measured telescopically, were spurious (Ingoli's and Kepler's anomalous remarks set aside). One such observation was the aforementioned daylight measurement of a star by Hooke. Also, Christian Huygens had reported in his 1659 *Systema Saturni* that the sizes of stars could be reduced to nothing by viewing them telescopically through increasingly darkened glass. And, in 1662 Johannes Hevelius published observations by Jeremiah Horrocks showing that stars wink out instantaneously when occulted by the moon (this in his *Mercurius in Sole Visus*, the same volume that contained Hevelius's own precise telescopic measurements of the apparent diameters of fixed stars).[51]

But such observations indicating the spurious nature of even telescopically measured star sizes came decades after Kepler. Moreover, the astronomical world would not quickly be convinced by these observations. For example, Huygens claimed that the telescope revealed the spurious nature of apparent star sizes and thus overthrew Brahe's star size argument, but John Flamsteed in 1702 argued directly against Huygens, stating that the body of Sirius seen through the telescope was as true as that of Mercury, and about the same apparent size, and stating that Huygen's claims were "plain prejudice". And as late as 1720 Edmund Halley was criticizing a newly published telescopic measurement of the diameter of Sirius and calculation of Sirius's physical bulk, showing it to exceed that of the orbit of the Earth.[52]

As long as stars were thought to have measurable diameters, Kepler's universe of giant dull stars and a unique, tiny, brilliant sun was the universe that science demanded. A universe in which stars were suns, such as was envisioned by Hooke or Bruno, did not stand up to observations. Thus there exists a phase in the Copernican Revolution, exemplified by Kepler, in which the Copernican universe was nearly as different from our modern view of the universe as was the geocentric universe. Indeed, that phase lasted up until about the point where Hooke discovered something about star sizes while trying to detect parallax.

FIGURES

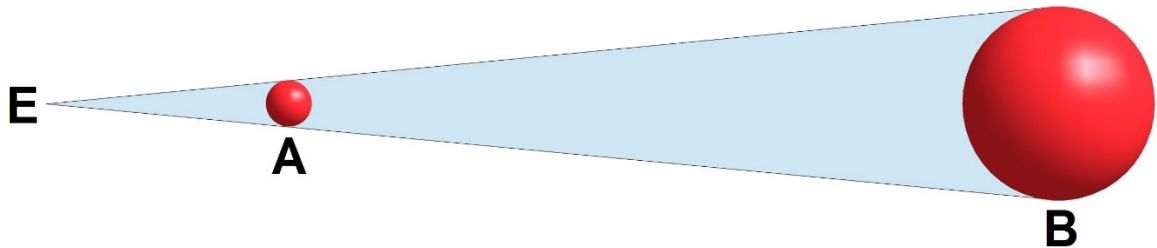

Figure. FS. The more distant an observed object is from Earth (E), the more bulky it must be to retain the same apparent size (indicated by the shaded region). A and B will each have the same apparent size as seen from E, but B is over four times more distant from E than is A, so B has over 4 times the diameter of A, and over 64 times (4 cubed) the volume of A.

C. M. Graney: The Starry Universe of Johannes Kepler, pre-print (page 19 of 28)

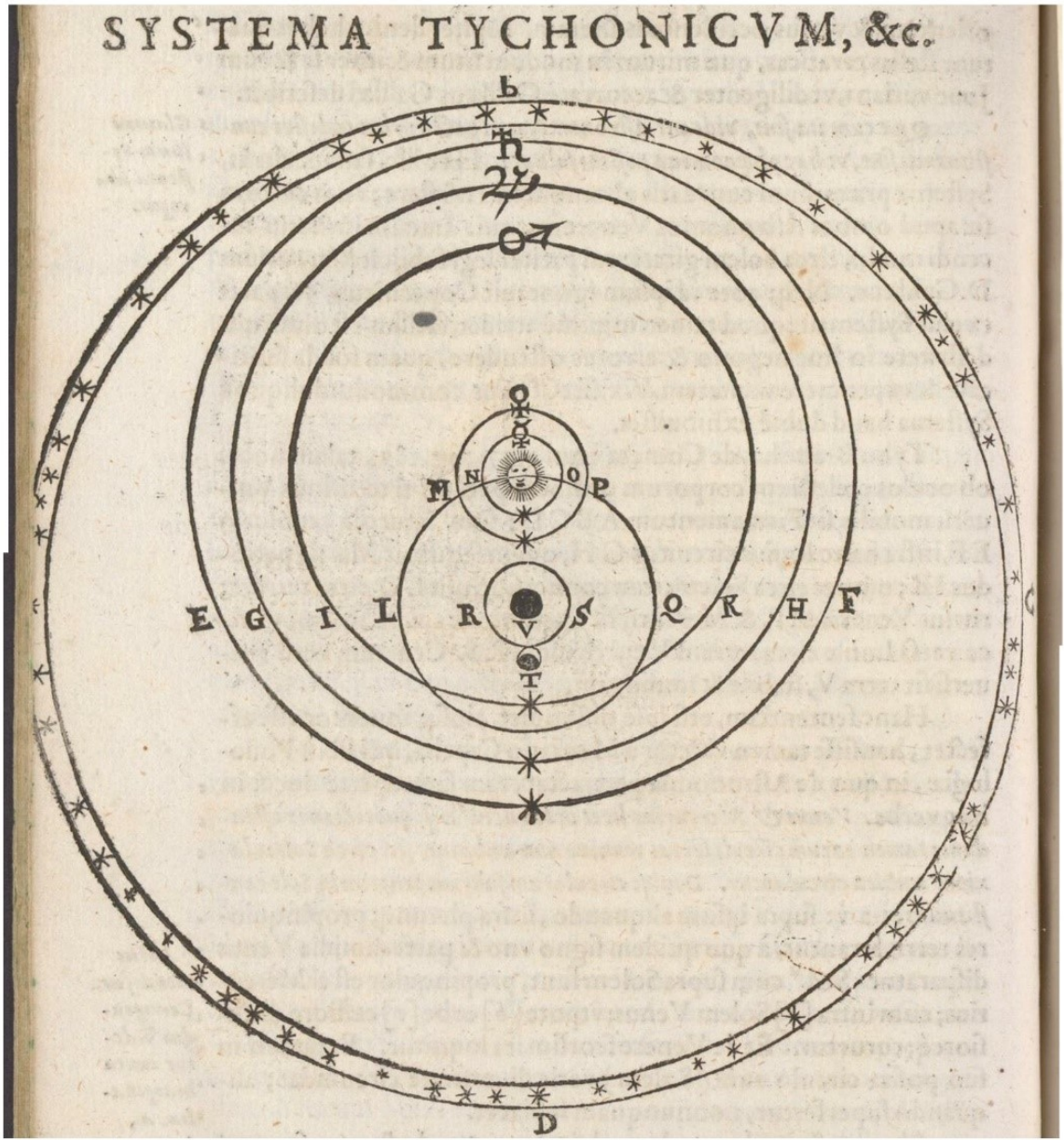

Figure TB. Tycho Brahe's hypothesis, as illustrated in Locher and Scheiner's *Disquisitiones Mathematicae*.[53] Earth is immobile at center. Mercury, Venus, Mars, Jupiter, and Saturn circle the sun as in the Copernican hypothesis, while the sun circles the Earth (as do the moon and stars). The fixed stars lie just beyond Saturn. Image credit: ETH-Bibliothek Zürich, Alte und Seltene Drucke.

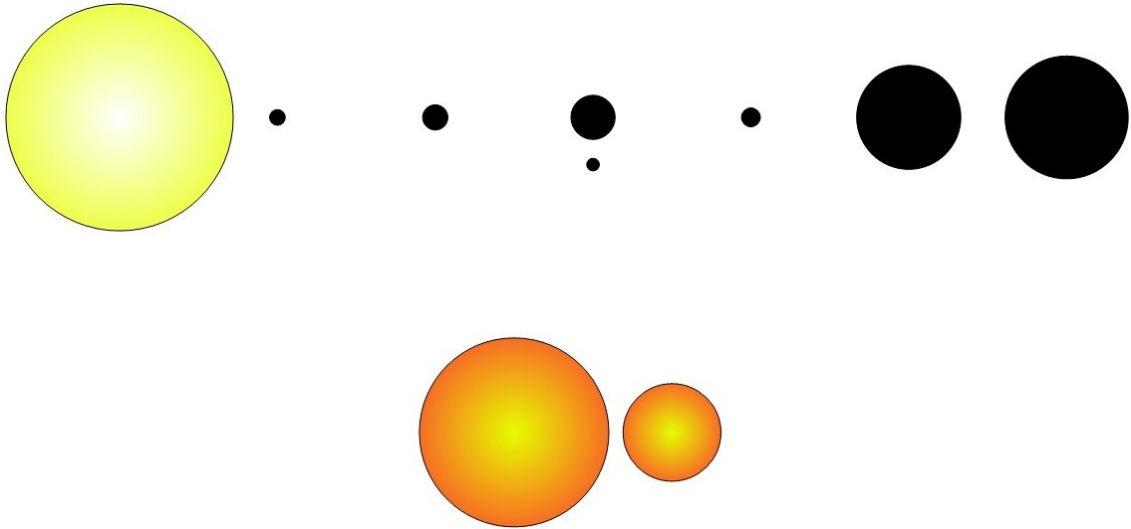

Figure TS. *Above*—the relative sizes of celestial bodies calculated by Tycho Brahe, based on his observations and measurements, for (from left to right, upper row) the sun, Mercury, Venus, Earth and moon, Mars, Jupiter, Saturn, as well as for (lower row) a large star and a mid-sized star in a hybrid geocentric universe (where the stars lie just beyond Saturn, as in Figure TB). Sun, stars, and planets all fall into a fairly consistent range of sizes. *Below*—the arrowed dots are the figure above, scaled to Brahe's calculated relative size for a mid-size star in the Copernican universe (where the stars lie at vast distances, and thus must be enormous to explain their apparent sizes as seen from Earth, as in Figure FS). Brahe said the huge Copernican stars were absurd.

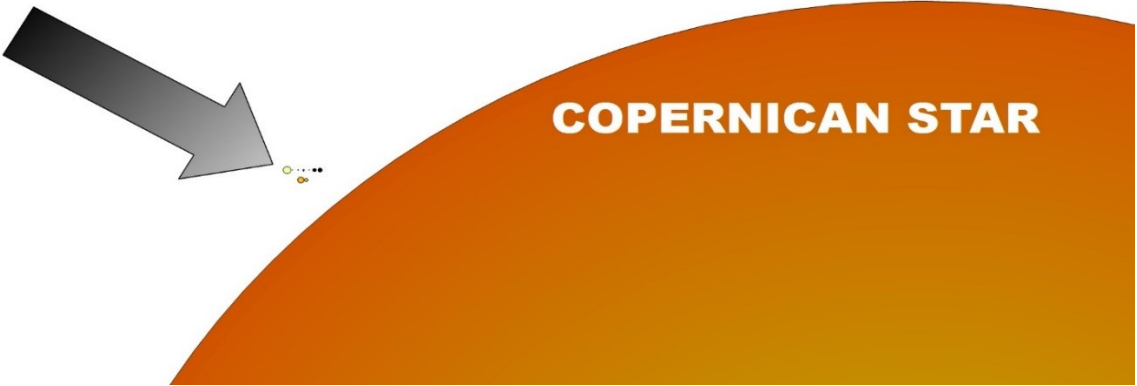

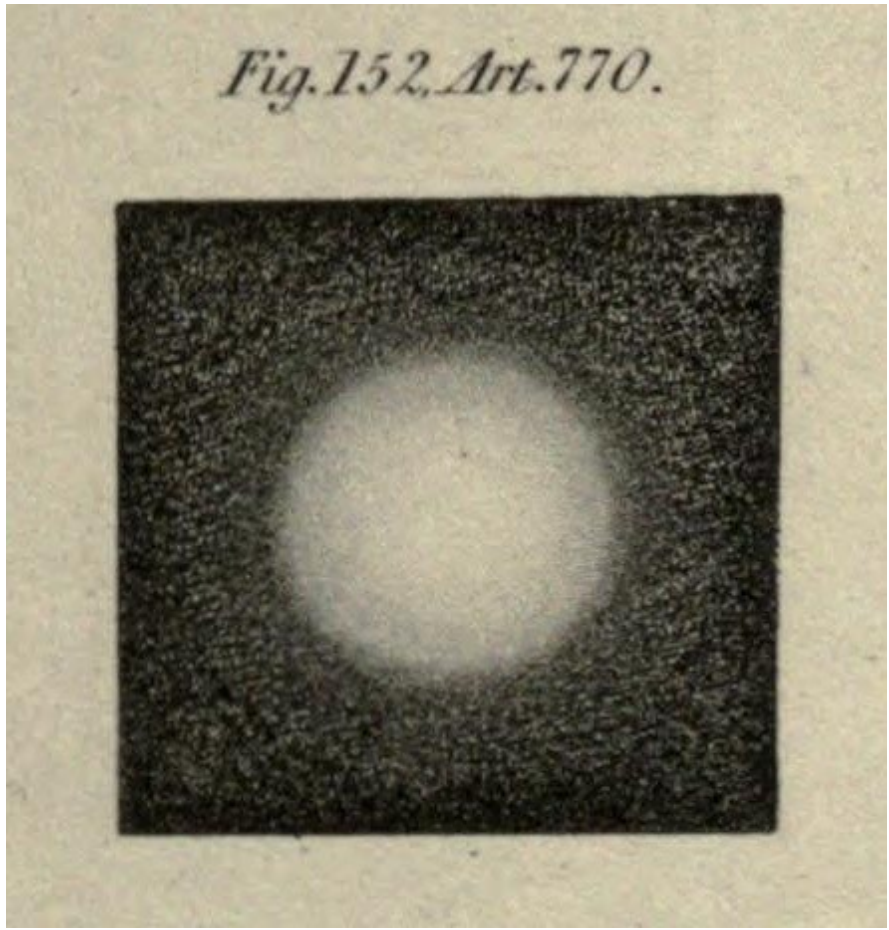

Figure D. A star as seen through a small aperture telescope, as illustrated in John F. W. Herschel's *Treatises on Physical Astronomy*.[54] This appearance of a sphere of measurable size is entirely spurious—an artifact of diffraction. However, early telescopic astronomers took such telescopic images to be the physical bodies of stars. Image credit: ETH-Bibliothek Zürich, Alte und Seltene Drucke.

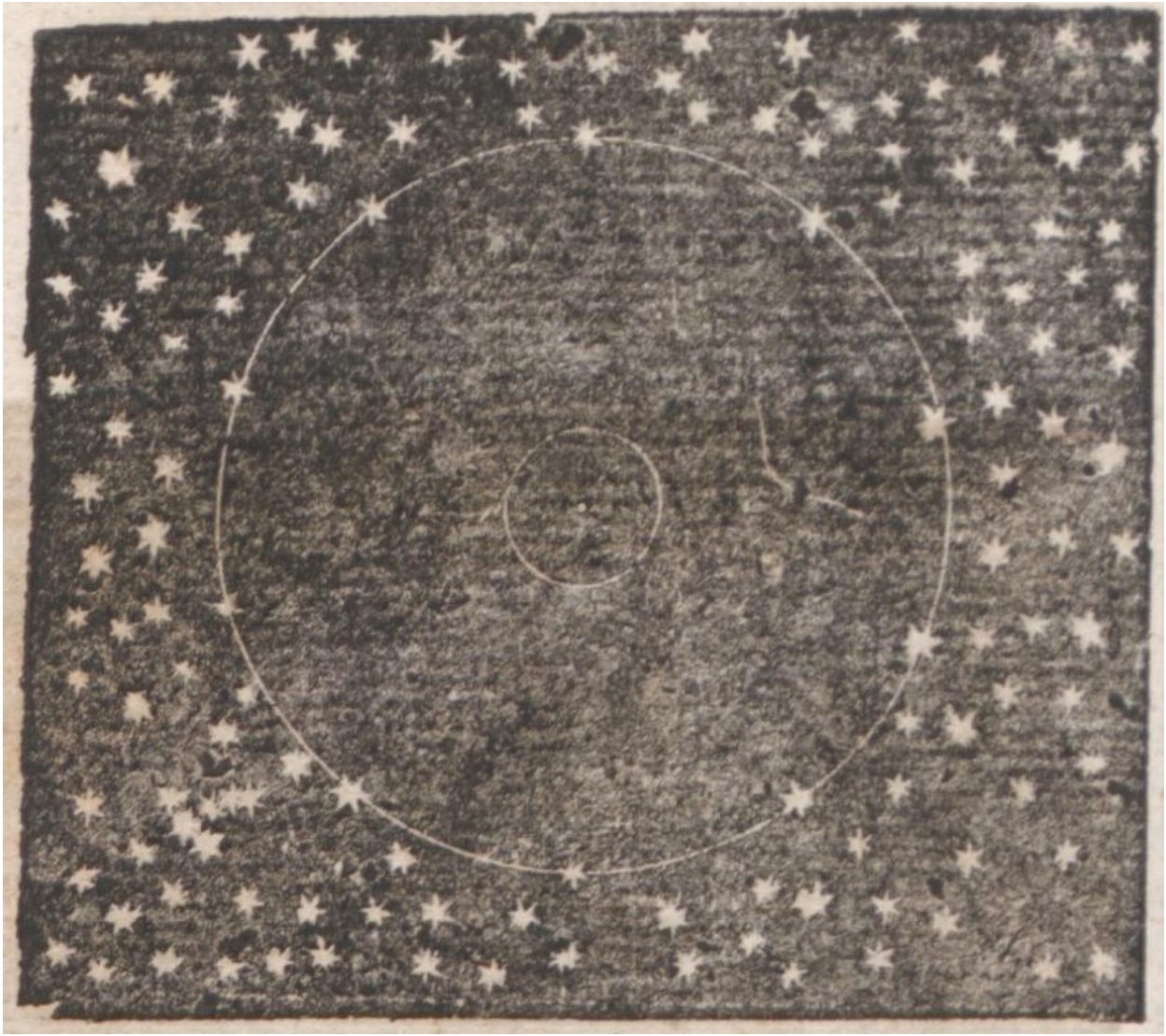

Figure EP.  Kepler's diagram from *Epitome*.[55]  The sun is the dot at the center.  Image credit: ETH-Bibliothek Zürich, Alte und Seltene Drucke.

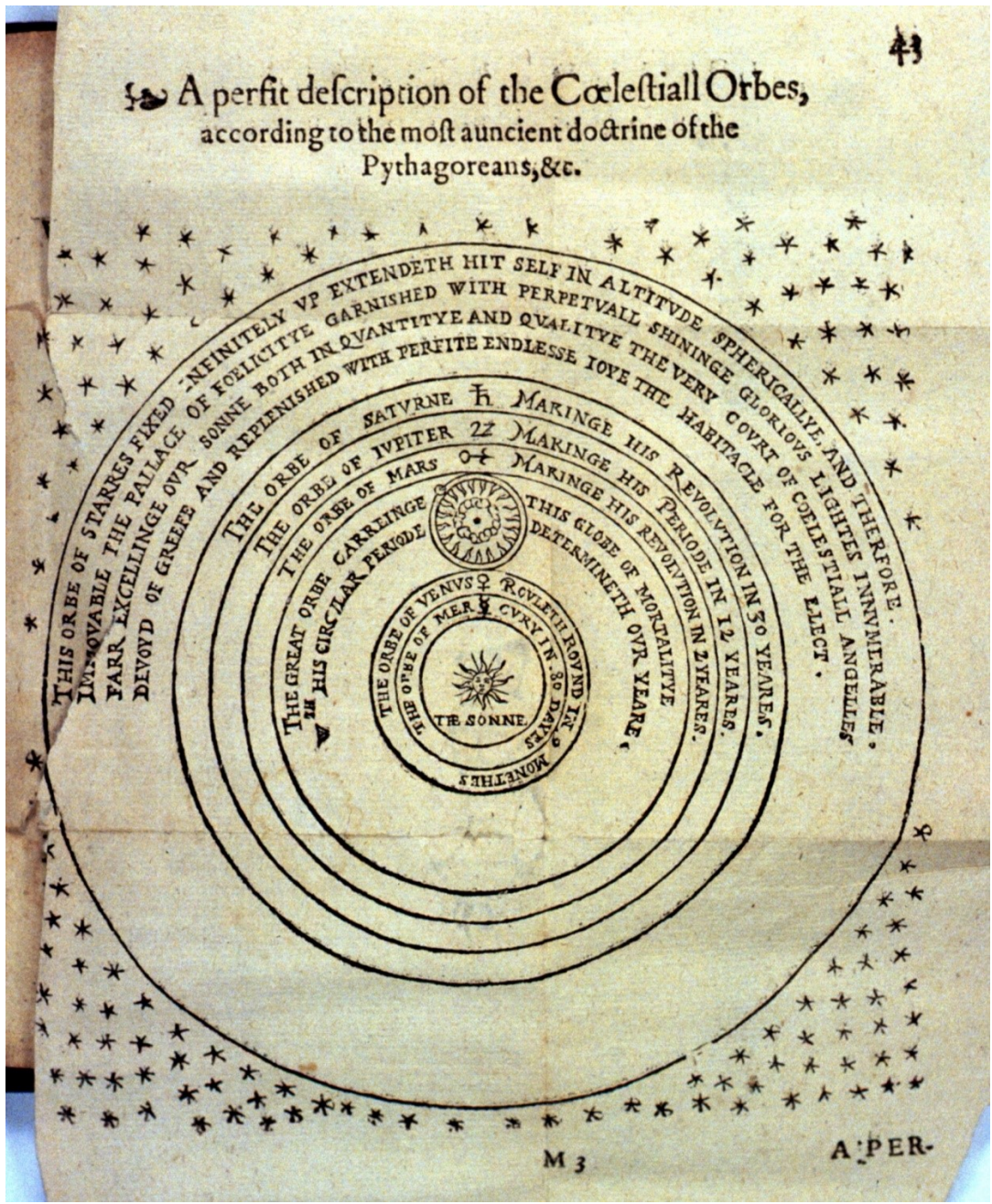

Figure TD. Sketch of the Copernican system by Thomas Digges. Note his description of the starry heavens as the "palace of felicity," with lights "far excelling our Sun both in quantity and quality," the court of the angels and the dwelling of the elect. From Leonard and Thomas Digges, *A Prognostication Everlasting* (London: 1576). Image(s) courtesy History of Science Collections, University of Oklahoma Libraries.

NOTES

[1] (Hooke 1674, 26)
[2] (Hooke 1674, 6, 8)
[3] (Hooke 1674, 6)
[4] (Van Helden 1985, 27, 30, 32, 50)
[5] (C. M. Graney 2017, 30)
[6] (C. Huygens 1722, 145)
[7] (Van Helden 1985, 51)
[8] For Marius, Crüger, Ingoli, and Riccioli, respectively, see (C. M. Graney 2015, 50-52, 52, 68-75, 103-139); for Mersenne, see (Mersenne 1627, 374-5).
[9] (C. M. Graney 2010)
[10] See (Graney and Grayson 2011) for a full discussion.
[11] (C. M. Graney 2015, 53-61)
[12] (C. M. Graney 2017, 29)
[13] (C. M. Graney 2015, 137)
[14] For secondary sources that discuss Chapter 16 see (Boner 2011, 101-106), (Van Helden 1985, 62-3), and (Westman 2011, 398-99). For a complete English translation of Chapter 16, see (Boner n.d.) or (C. Graney 2018)—the latter source is a pre-print and has not been peer reviewed.
[15] (Kepler 1606, 83): "Braheus... dum concinnitatem in perfectissimo opere desiderat; si Sphaerae unius fixarum tam insana sit vastitas; mobilium vero omnium tam contempta exilitas. Quemadmodum, ait, in corpore humano ingens vitium, si digitus, si nasus, multis partibus superet molem totius reliqui corporis."
[16] (Kepler 1606, 87-88)
[17] (Kepler 1606, 86-87); Kepler obtains the 34,000,000 figure thusly: the Earth-sun distance is 1,432 semidiameters of Earth; the sun-Saturn distance is ten times that, or 14,320; the sun is six Earth semidiameters, so the sun-Saturn distance is 14,320 / 6 ≈ 2387 times the solar semidiameter; thus the sun-stars distance is 2387 times the sun-Saturn distance of 14,320 Earth semidiameters, or 2387 × 14,320 ≈ 34,000,000 Earth semidiameters.
[18] (Kepler 1606, 86): "Mundi perfectio est motus...."
[19] (Kepler 1606, 83): "Pulchra proportio, ubi semper velocior, qui est Soli quiescenti, motusque omnis dispensatori propinquior."
[20] 20,000 × π = 63,000. As circumference is π times diameter rather than semidiameter, this number is too small by half. There are a variety of such typos in Kepler's text of Chapter 16.
[21] 63,000 / 24 = 2625
[22] 860 × 2625 = 2,257,500
[23] 2,257,500 / 300 = 7,525
[24] (Kepler 1606, 84): "Ito nunc ad Ptolemaeum, & antiquam sententiam; omnia invenies incredibiliora. In illa semidiameter Sphaere fixarum vicies millenas Telluris semidiametros possidet; ambitus igitur erit sexagies ter millium. Modesta sane multitudo, comparata ad Copernicanam; sed quae omnis in uno die circumire dicitur. Debentur igitur uni horae semidiametri 2625: quarum quaelibet 860 milliaria continet. Hic vide mihi immensum discrimen; Saturnus, qui est apud Ptolemeum fixis proximus, ut eas tantum non tangat, Copernico in una hora trajicit per 300 milliaria, Ptolemaeo vicies bis centena millia quinquagies septies mille quingenta milliaria. Credendus est igitur velocior apud Ptolomaeum, quam est apud Copernicum, septies mille, quingenties vicies quinquies. Quicunque tentaverit mente comprehendere hanc incredibilem velocitatem; aeque fatigatur, & vehementius etiam, quam qui Copernicanam immensitatem."
[25] (Kepler 1606, 86)
[26] (Kepler 1606, 85)
[27] (Kepler 1606, 88): "Ubi superat magnitudo, ibi deficit perfectio, & in molis deminutionem succedit nobilitas. Amplissima sane est Copernico Sphaera fixarum; sed iners, motu nullo. Sequitur Mundus mobilis. Hic jam quanto minor tanto divinior quod motum accepit tam admirabilem, tam ordinatum. Neque tamen vegetante facultate

constat locus iste; neque ratiocinatur, neque discurrit: quod agit (dum movetur,) non didicit, sed impressum sibi a principio retinet; quod non est, neque erit unquam; quod est, id a seipso non est factus; idem manet, qui conditus est. Succedit ergo pilula haec nostra, tuguriolum nostrum; quod Tellurem dicimus, matrix vegetabilium, ipsa intus informata facultate quadam, mirabilium operum architectatrice; quae accendit de se ipsa tot stirpium, tot piscium, tot insectorum animulas quotidie; ut facile molem reliquamprae hac sua nobilitate contemnat. Denique vide mihi corpuscula, quae animalia dicimus, quibus quid exilius in comparatione mundi fingi potest? At ibi jam sensus, & voluntarij motus, architectura corporum infinita. Vide mihi inter illa, pulvisculos hos, quos Homines dicunt; quibus Creator hoc dedit, ut quodammodo a seipsis nascantur, seipsos vestiant, arment, doceant infinitas artes, & quotidie proficiant in melius; in quibus Dei imago; qui domini quodammodo sunt totius molis. Et quis est nostrum, qui optet sibi corpus, Mundi amplitudine, ut pro ea careat anima? Discamus igitur creatoris bene placitum; qui & rudis molis, & minutorum perfectionis author est: nec tamen mole gloriatur, sed nobilitat illa, quae minuta esse voluit.  Denique per haec intervalla a Tellure ad Solem, a Sole ad Saturnum, a Saturno ad fixas, discamus paulatim conscendere ad agnoscendum divinae potentiae immensitatem."

[28] (Kepler 1606, 89): "Et haec, de objecta Copernico vastitate fixarum, tanto libentius inserui, quod pertinuerint ad incredibilem novi sideris magnitudinem aestimandam. Nam si quatuor solum minuta occupavit (quantus Sirius apparet) jam per hanc hypothesin Copernici tota machina mobilium multo fuit major; ut cui tria solum minuta tribuebamus supra, si quis illam a fixis respiceret."

[29] (Kepler 1891, 175): "quaero ad quas leges examinet opera manuum Dei, ut ea improportionata dicat. Ostendi maiorem esse proportionem inter scirum animalculum subcutaneum in manu hominis, et serpentem illum africanum.... cur nimium est in oculis Ingoli quod continent 16506000 semidiametros Terrae, nec est nimium quod 14000 continetur? quid simile faciunt homines? quibus exemplis humanis confirmatus Ingoli animus repudiet opera Dei ut nimia?"

[30] (C. M. Graney 2015, 68-73).  A complete English translation of Ingoli's essay is available in Appendix A of this source.

[31] (Kepler 1891, 175): "Convincit, inquit, fixas nihil in terram operari? nihil de operatione dicamus, re non ab omnibus confessa: dicamus de illuminatione, quae est operatio quae patet oculis, cur quae per 14000 semidiametros illuminant, non illuminent per 16506000? Si millies ducenties sunt remotiores, erunt et toties maiores: ita effectus illuminationis terrae manebit idem."  Italics added.

[32] (Kepler 1891, 175): "de stella nova... ubi pluribus dissolvi praetensam absurditatem magnitudinis fixarum"

[33] Also see (Granada 2008) on Kepler and Bruno in Kepler's *De Stella Nova*, *Dissertatio cum Nuncio Sidereo*, and *Epitome Astronomiae Copernicanae*.

[34] (Koyré 1957, 69)

[35] (Van Helden 1985, 63-64)

[36] As one 2' star located on a sphere of stars 81' from a second star on that sphere would subtend an angle of 2 / 81 = 0.0247 = 1.4° (or 85') as seen from that second star, Kepler's value of 2.75° (≈ 2.8°) is double what it should be.

[37] (Kepler 1606, 106); translation from (Koyré 1957, 63).

[38] (Koyré 1957, 64-70)

[39] (Kepler 1606, 108): "Nam si sic res habeat, uti dictum; Certe ut quaelibet duplo, triplo, centuplo altior, ita duplo, triplo centuplo erit major.  Quippe quantumcunque dicas elevatam; nunquam efficies, ut non videatur habere a nobis duum minutorum diametrum."  For a similar translation, see (Koyré 1957, 68), but Koyré translates the double negative "nunquam... non" as a single, and thus says that the star will never reach two minutes in size.

[40] (Kepler 1965, 34-35), italics added.

[41] (Kepler 1965, 34)

[42] (Kepler 1965, 35-36)

[43] (Koyré 1957, 81), (Kepler 1618, 37); here Kepler corrects the size error mentioned in an earlier note regarding the *De Stella Nova* discussion of Orion belt stars; in the *Epitome* discussion he states that the size of one star seen from another will be just under 85'.

[44] (Koyré 1957, 84-85), (Kepler 1618, 39)

[45] (Koyré 1957, 76-86)

[46] (Kepler 1618, 36)

[47] (Van Helden 1985, 87-89)

⁴⁸ (C. M. Graney 2010)—Galileo, Riccioli, Hevelius; (C. M. Graney 2009)—Hevelius; (C. M. Graney 2015, 45-61)—general discussion.

⁴⁹ (Kepler 1606, 106), paraphrasing the following paragraph:

> Sed quia secta haec abutitur authoritate Copernicanae adeoque universae Astronomiae; quod Copernicus fixas immobiles, omnis vero Astronomia, ac praecipue Copernicana, incredibiliter altas praestet: Age petamus etiam ab ipsa Astronomia remedium: Ut, cujus indulgentia proritata isthaec philosophantium insania, ruptis locis & repagulis, sese in hanc immensitatem extulit: ejusdem etiam artibus & blandientibus poppysmis revertatur intra Mundi metas, atque carceres suos. Certe equidem vaganti per illud infinitum bene non est.

Koyré (p. 61) deals with this paragraph through a combination of translation (indicated here with added italics) and paraphrase which does not suggest insanity:

> *But because this sect misuses the authority of Copernicus as well as that of astronomy in general, which prove—particularly the Copernican one—that the fixed stars are at an incredible altitude: well then we will seek the remedy in astronomy itself.*
>
> Thus by the same means which seem to those philosophers to enable them to break out of the limits of the world into the immensity of infinite space, we will bring them back. *"It is not good for the wanderer to stray in that infinity."*

Miguel Granada (Granada 2008, 479), however, translates this as referring to the madness of those philosophers who—

> misuse the authority of Copernicus as well as that of astronomy in general, which proves—particularly that of Copernicus—that the fixed stars are at an incredible altitude. Well, let us seek the remedy in Astronomy herself, so that by her arts and soothing blandishments this madness of the philosophers (a madness that was provoked by her indulgence, once the bolts were broken and confined spaces abandoned [*ruptis locis et repagulis*] and carried itself out into this immensity), might be led to come back within the bounds of the world and its prisons [*intra Mundi metas, atque carceres suos*]. Surely, it is not good to wander through that infinity [*vaganti per illud infinitum bene non est*].

The words Kepler uses in the portion that Koyré paraphrases—*insania*, *repagulis* (the bolts or bars of a door), *blandientibus poppysmis* (alluring cluckings of the tongue, like might be used to call back a pet), *carceres* (confinement cells)—all convey a tone that suggests dealing with the insane. A still closer translation of this paragraph:

> But because this sect abuses the authority of Copernican, and indeed all, Astronomy; insofar as Copernicus presents as incredibly high the unmoving fixed stars (true of all Astronomy, but especially Copernican): Come! Indeed we may seek from Astronomy herself the remedy: when that insanity of the philosophizers, provoked by her indulgence, has carried itself away into this immensity, quarters and door-bars destroyed: it may still be turned back within the boundaries of the universe, and [they within] their own cells, by her alluring arts and tongue-cluckings. Certainly it is not good, truly wandering through that infinity.

Put briefly, Kepler is saying that the crazy philosophers have broken out of the asylum and are running around in the infinity they have created for themselves, but through astronomy we can gently coax them back into their cells for their own good.

⁵⁰ This discussion on Digges, Rothmann, and Lansbergen is condensed from (C. M. Graney 2015, 76-85); (C. M. Graney 2013).

⁵¹ The question is worth further investigation of whether observations in the later seventeenth century such as those published by Huygens, Hevelius, and Hooke are in fact the earliest solid, reproducible evidence for the spurious nature of telescopically observed stellar disks, and in particular whether the anomalous remarks about stars by Ingoli and Kepler previously mentioned might be more than just anomalous. Some times Galileo's remarks in his 1610 *Sidereus Nuncius* regarding fixed stars seen through a telescope not having a circular periphery are taken as him recognizing the spurious nature of the telescopic appearance of stars—e.g Stillman Drake in (Galilei 1957, 47, n. 16). However, Galileo had only been using a telescope a short time when he wrote those remarks. His subsequent discussions of the telescopic appearance of stars over many years consistently refer to stellar disks, per Figure D. See (Graney and Grayson 2011); (C. M. Graney 2015, 45-61).